# Oscillator Circuit for Spike Neural Network with Sigmoid Like Activation Function and Firing Rate Coding

**Andrei Velichko [1]*, Petr Boriskov [1]**

[1] Institute of Physics and Technology, Petrozavodsk State University, 31 Lenina str., Petrozavodsk 185910, Russia; rectorat@petrsu.ru

* Correspondence: velichko@petrsu.ru; Tel.: +79114005773

**Abstract:** The study presents an oscillator circuit for a spike neural network with the possibility of firing rate coding and sigmoid-like activation function. The circuit contains a switching element with an S-shaped current-voltage characteristic and two capacitors; one of the capacitors is shunted by a control resistor. The circuit is characterised by a strong dependence of the frequency of relaxation oscillations on the magnitude of the control resistor. The dependence has a sigmoid-like form and we present an analytical method for dependence calculation. Finally, we describe the concept of the spike neural network architecture with firing rate coding based on the presented circuit for creating neuromorphic devices and artificial intelligence.

**Keywords:** firing rate coding; spiking neuron; neuromorphic; oscillator; non silicon neuron

## 1. Introduction

Spike-based models are extremely popular for the development of artificial neural network architectures [1]. Spike neural networks (SNNs), also called third generation neural networks [2], functionally resemble the operation of real biological neurons. In 1926, E.D. Adrian [3] established the dependence of the frequency of spike generation on the impact strength on biological sensory nerve fibres. Similar dependences were observed for neural receptors that respond to influences of a very different nature: mechanoreceptors (compression, stretching), changes in temperature, lighting, etc. [3–6]. This led to the development of two main methods of encoding information in SNN: temporal coding and firing rate coding [7]. The temporal coding method encodes information with time intervals between individual spikes, or with effects of spike synchronization [8,9]. The firing rate coding method encodes information in changing the spike repetition rate, measures the number of spikes in a certain time window and can have two similar strategies: spike-count code and rate code [8]. In the spike-count model, a neuron generates a postsynaptic spike if a certain number of presynaptic spikes accumulate at its input. In the rate code model, the frequency of generation of output spikes depends on the number of neurons arriving for a certain period.

A number of models of spike neurons exists in the literature, including the Hodgkin and Huxley model with a high degree of bio-similarity [10], simplified Izhikevich neuron models [11], FitzHugh-Nagumo and FitzHugh-Rinzel models [12], and the very simple and most popular leaky integrated-and-fire (LIF) neuron model with threshold activation function.

The analog principle of brain functioning is fundamentally different from digital implementation, which stimulated the development of neuromorphic engineering [13,14], which develops circuit solutions of known neural models. Two directions can be distinguished: the creation of silicon neurons (SiNs) circuits and the development of neural circuits with non-silicon technology, based on materials with significant physical effects, for example, $VO_2$ with a metal insulator transition [15]. SiNs are implemented on analog / digital very large scale integration (VLSI) circuits that emulate the electrophysiological behavior of real neurons [14].

The development of non-silicon neurons technology opens the possibility of creating compact circuits with a small number of elements. For example, a LIF neuron can be created on a simple circuit of a relaxation generator with a switching element with an S-shaped I–V characteristic, also called a memristor [12,15–17]. A wide range of nonlinear elements with an S-shaped I – V characteristic, which contain sections with negative differential resistance (NDR), is available in modern electronics. The NDR effect and the related electrical instability are caused by the presence inside the element of either positive current feedback (S-type) or voltage feedback (N-type) [18–21]. A traditional element with an S-type I – V characteristic is a silicon dynistor (trigger diode), and transition metal oxides exhibit this characteristic among non-silicon materials [21]. In niobium and vanadium dioxides, electrical instability is caused by the metal-insulator phase transition (MIT), when heating the material by the flowing current (at a certain temperature) leads to the electronic-structural transition of the oxide from the semiconductor state to the metallic state [22].

The literature describes the spike-frequency adaptation mechanisms in the LIF neuron, which are based on variations in the supply current of the neuron circuit [14,16], or changes in the spike activation threshold, for example, threshold voltage [14]. The threshold voltage of the S-shaped I – V characteristic of a non-silicon neuron is more difficult to control than the current in the circuit; nevertheless, this can be done by varying, for example, the temperature of the switching $VO_2$ element [23].

In the current study, we present an effective way to control the frequency of a relaxation oscillator (LIF neuron) by varying a resistance of a control variable resistor. We found a strong dependence of the frequency of the output relaxation oscillations on the control resistance, which has an explicit sigmoidal shape. The concept of using this circuit in spike neural networks by implementing firing rate coding is proposed.

## 2. Oscillator circuit and methodology for calculating the oscillation frequency

The basic element of the oscillator circuit is an electric switch with an S-shaped current-voltage characteristic, presented in Figure 1a, where the current $I_{sw}$ is controlled by the applied voltage $U_{sw}$ according to the law:

$$I_{sw}(U_{sw}) \approx \begin{cases} U_{sw}/R_{off}, & \text{state} = \text{OFF} \\ U_{sw}/R_{on}, & \text{state} = \text{ON} \end{cases} \quad (1),$$

Switching in (1) between OFF and ON states happens in accordance with the following conditions:

$$\text{state} = \begin{cases} \text{OFF}, & \text{if (state = ON) and } (U < U_h) \\ \text{ON}, & \text{if (state = OFF) and } (U > U_{th}) \end{cases} \quad (2).$$

Parallel connection to the switch of a capacitance ($C_0$) and setting the supply current of the circuit in the NDR range

$$I_{th} < I_0 < I_h \quad (3),$$

implement a typical relaxation oscillator circuit (Figure 1b). This is a basic LIF neuron circuit, and, for S-type I-V characteristic (1-2), it has simple analytical dependences of the current (or voltage) on the switch from time. Figure 1c presents an example of self-oscillations oscillograms of current and voltage on the switch, with parameters $U_{th}$=4 V, $U_h$=2 V, $R_{off}$=40 kΩ and $R_{on}$=200 Ω, $I_0$=150 μA, $C_0$=1 μF.

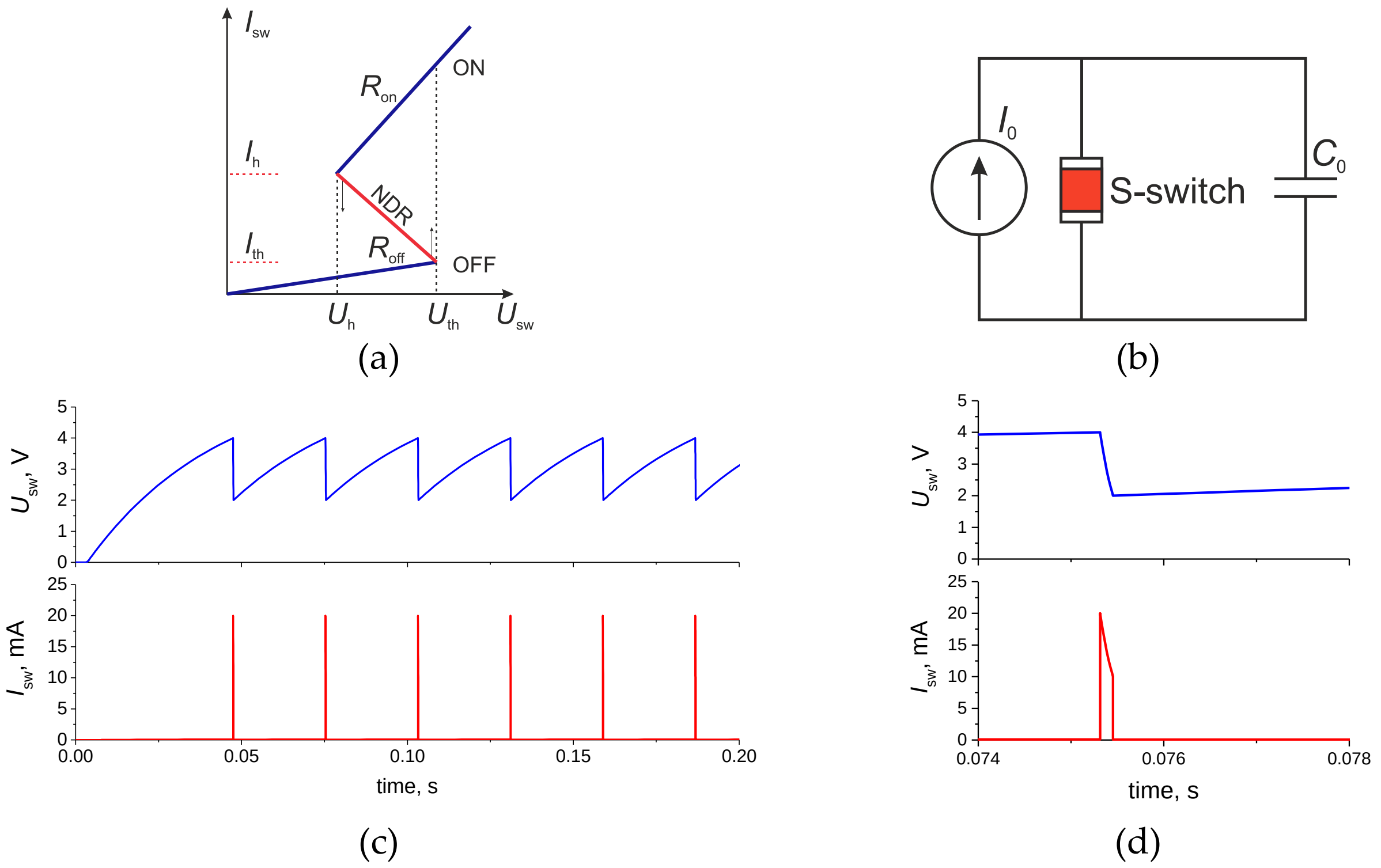

**Figure 1.** Schematic representation of the S-shaped I-V characteristic of the switch, indicating the location of the threshold parameters and NDR (a). The basic circuit of the relaxation oscillator (b) and voltage and current oscillations (c, d) at $U_{th}$=4 V, $U_h$=2 V, $R_{off}$=40 kΩ and $R_{on}$=200 Ω, $I_0$=150 μA, $C_0$=1 μF. The simulation was performed in the LTspice software.

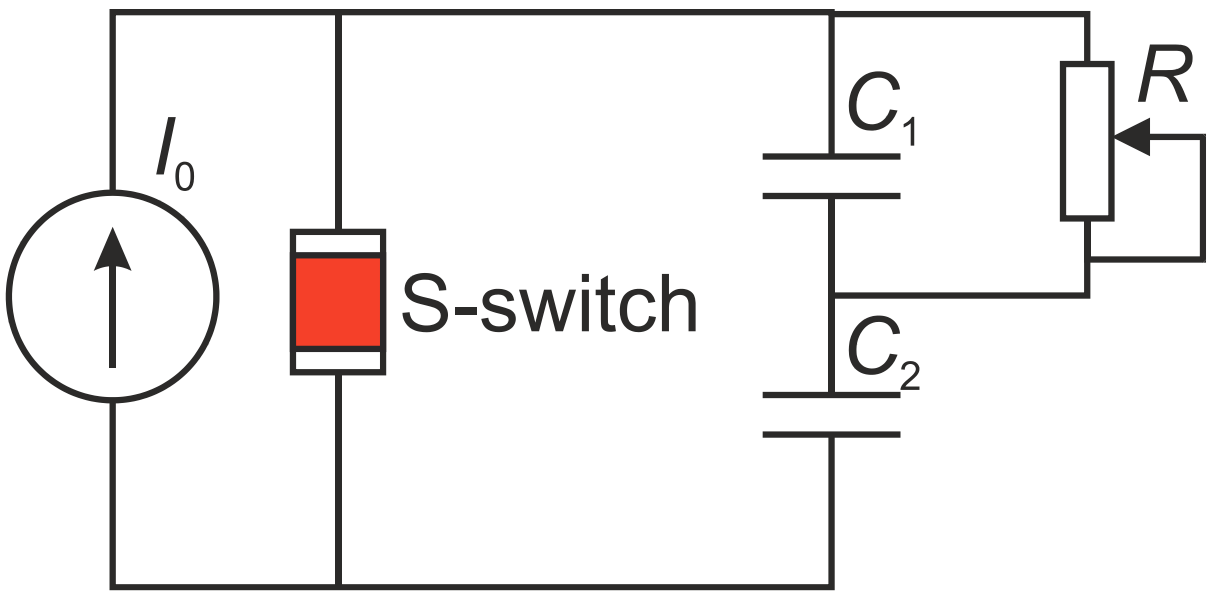

**Figure 2.** Oscillator circuit with a variable resistor R.

As Figure 1d demonstrates, the voltage oscillations occur in the range between $U_h$=2 V and $U_{th}$=4 V. At the time of switching, the current increases sharply, and then the current slowly decreases, when the capacitor $C_0$ is discharged.

The circuit under study is presented in Figure 2, where, instead of the capacitance $C_0$, two serial capacitors are installed ($C_1$ and $C_2$), and one of the capacitors is connected in parallel with a variable resistor $R$. Example of voltage oscillograms on the switch $U_{sw}(t)$ and capacitors $U_1(t)$, $U_2(t)$ are demonstrated in Figure 3a. The current oscillogram is shown in Figure 3b, with parameters $U_{th}$=4 V, $U_h$=2 V, $R_{off}$=40 kΩ and $R_{on}$=200 Ω, $I_0$=3 mA, $C_1$=10 nF, $C_2$=1 μF, $R$=100 Ω.

The voltage oscillogram on the switch $U_{sw}(t)$ has a more complex form than in the basic circuit (see Figure 1b). To understand the functioning of the new circuit, we present the derivation of the analytical dependences $U_{sw}$(t), $U_1(t)$ and $U_2(t)$.

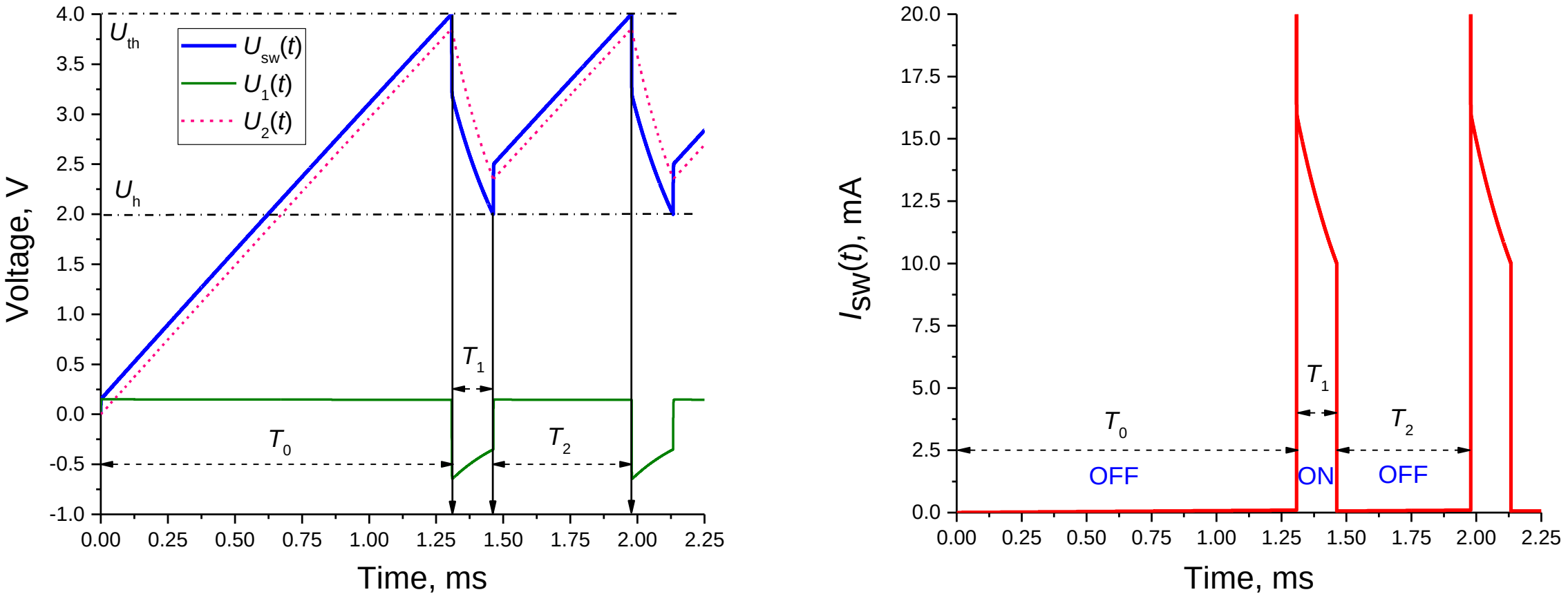

(a) (b)

**Figure 3.** Voltage oscillograms on switch $U_{sw}(t)$ and capacitors $U_1(t)$, $U_2(t)$ of the studied circuit (a), current oscillogram of switch $I_{sw}$(t) of the studied circuit (b). $T_0$ is the time until the switch is first turned on, $T_1$ is the duration of the ON state of the switch, $T_2$ is the duration of the OFF state of the switch. Oscillation period $T=T_1+T_2$. Circuit parameters: $U_{th}$=4 V, $U_h$=2 V, $R_{off}$=40 kΩ and $R_{on}$=200 Ω, $I_0$=3 mA, $C_1$=10 nF, $C_2$=1 μF, $R$=100 Ω. The simulation was performed in the LTspice software.

Based on the Kirchhoff equations, we obtain a system of differential equations for capacitor voltages ($U_1(t)$ and $U_2(t)$):

$$\begin{cases} \dfrac{\mathrm{d}U_1}{\mathrm{d}t} = a_1\left(V_0 - bU_1 - U_2\right) \\ \dfrac{\mathrm{d}U_2}{\mathrm{d}t} = a_2\left(V_0 - U_1 - U_2\right) \end{cases} \quad (4),$$

where the parameters are equal to

$$a_1 = \frac{1}{R_{sw}C_1},\ a_2 = \frac{1}{R_{sw}C_2},\ b = \frac{R+R_{sw}}{R}, V_0 = R_{sw}I_0 \quad (5),$$

and the switch resistance is $R_{sw}$ = $R_{on}$ or $R_{off}$ in the ON and OFF states, respectively.

Solution (4) has the following general form (with arbitrary coefficients $M_1$ and $M_2$)

$$\begin{cases} U_1(t) = -a_1\left(M_1 \cdot \exp(-\alpha_1 t) + M_2 \cdot \exp(-\alpha_2 t)\right) \\ U_2(t) = V_0 + M_1\beta_1 \cdot \exp(-\alpha_1 t) + M_2\beta_2 \cdot \exp(-\alpha_2 t) \end{cases} \quad (6),$$

where the parameters of the homogeneous system of equations, associated with (4), are calculated as

$$\alpha_1 = \frac{a_1 b + a_2 - \sqrt{D}}{2},\ \alpha_2 = \frac{a_1 b + a_2 + \sqrt{D}}{2},$$
$$\beta_1 = \frac{a_1 b - a_2 + \sqrt{D}}{2},\ \beta_2 = \frac{a_1 b - a_2 - \sqrt{D}}{2}, D = 4a_1a_2 + \left(a_1 b - a_2\right)^2 \quad (7).$$

Knowing the initial values $U_1(t=0)=U_{10}$ and $U_2(t=0)=U_{20}$, we can find the coefficients $M_1$ and $M_2$ of solution (6):

$$M_1 = \frac{a_1(U_{20} - V_0) + U_{10}\beta_2}{a_1(\beta_1 - \beta_2)},\ M_2 = -\left(\frac{a_1(U_{20} - V_0) + U_{10}\beta_2}{a_1(\beta_1 - \beta_2)} + \frac{U_{10}}{a_1}\right) \quad (8).$$

In the mode of periodic relaxation oscillations, the values of $U_{10}$ and $U_{20}$ are not known. To find them, it is necessary to start the calculation from the moment the supply current $I_0$ is turned on, when the voltage across the capacitors is known ($U_{10}=0$ and $U_{20}=0$), and the switch has the resistance $R_{sw} = R_{off}$. In this way, it is possible to obtain a solution to the desired voltage on the switch $U_{sw}^{(0)}(t)=U_1^{(0)}(t)+U_2^{(0)}(t)$. The time $t=T_0$ can be found (see Figure 3), when the first switching occurs, that is $U_{sw}^{(0)}(T_0)=U_{th}$, and solution (6) proceeds to the mode with $R_{sw} = R_{on}$. The initial values of $U_{10}$ and $U_{20}$ are now determined by the final values of $U_1^{(0)}(t=T_0)$ and $U_2^{(0)}(t=T_0)$, and they are equal to the voltages that the capacitors $C_1$ and $C_2$ have at the time of the first switching $t=T_0$.

From this point on, the circuit behaves periodically. The first periodic solution describes the process of decreasing $U_{sw}^{(1)}(t)= U_1^{(1)}(t)+U_2^{(1)}(t)$ to the threshold voltage $U_h$ that will occur at time $t=T_1$ (see Figure 3), and the switch goes into OFF state with $R_{sw} = R_{off}$. The initial values for the second periodic solution $U_{10}$ and $U_{20}$ are now equal to the final values $U_1^{(1)}(t=T_1)$ and $U_2^{(1)}(t=T_1)$ of the first periodic solution (6). Finally, at some point in time $t=T_2$, the switch goes into the ON state, and the process will be repeated between the first and second solutions (6). The period of relaxation oscillations is equal to the sum of the periods of the on and off states of the switch $T=T_1+T_2$, and the frequency is expressed as

$$F = \frac{1}{T_1 + T_2} \quad (9).$$

The times ($T_0$, $T_1$, $T_2$) are the solutions of transcendental equations:

$$U_1^{(0)}(T_0) + U_2^{(0)}(T_0) = U_{th},\ U_1^{(1)}(T_1) + U_2^{(1)}(T_1) = U_h,\ U_1^{(2)}(T_2) + U_2^{(2)}(T_2) = U_{th} \quad (10),$$

where the superscripts (0, 1, 2) indicate the solution before the first switch on, and the first and second periodic solutions (6), respectively. In Equations (4-10), the function $b(R)$ and its associated parameters $\alpha_1(R)$, $\alpha_2(R)$, $\beta_1(R)$, $\beta_2(R)$, $D(R)$ depend on the resistance $R$ in (5).

## 3. The results of the circuit study

With an uniform increase with time in the resistance $R$ from 0 to 300 Ω, and with the circuit parameters $I_0$=150 μA, $C_1$=10 nF, $C_2$=1 μF, we observe an intense change in the oscillation frequency $F$ depicted in Figure 4.

Figure 5a presents the dependences of the oscillations frequency $F(R)$ of the circuit (Figure 2) calculated for various capacitances $C_1$. With growing $R$, the frequency sharply increases in a certain range of resistances $R$ ~150-200 Ω, and then it gradually (with a less steep slope) reaches a constant value $F(R\to\infty) = F_{max}$. The greater the ratio $C_2/C_1$, the greater is the frequency jump $F_{max}/F(0)$, which can reach several orders of magnitude. For small $R$, when the capacitance $C_1$ is shunted by the resistance, the oscillation frequency is determined by the bigger capacitance ($C_2$) and the frequency has small values (tens of Hz). For large values of resistance $R$, the circuit operates with two successive capacitances, where a small capacitance ($C_1$) is dominant and determines a high oscillation frequency.

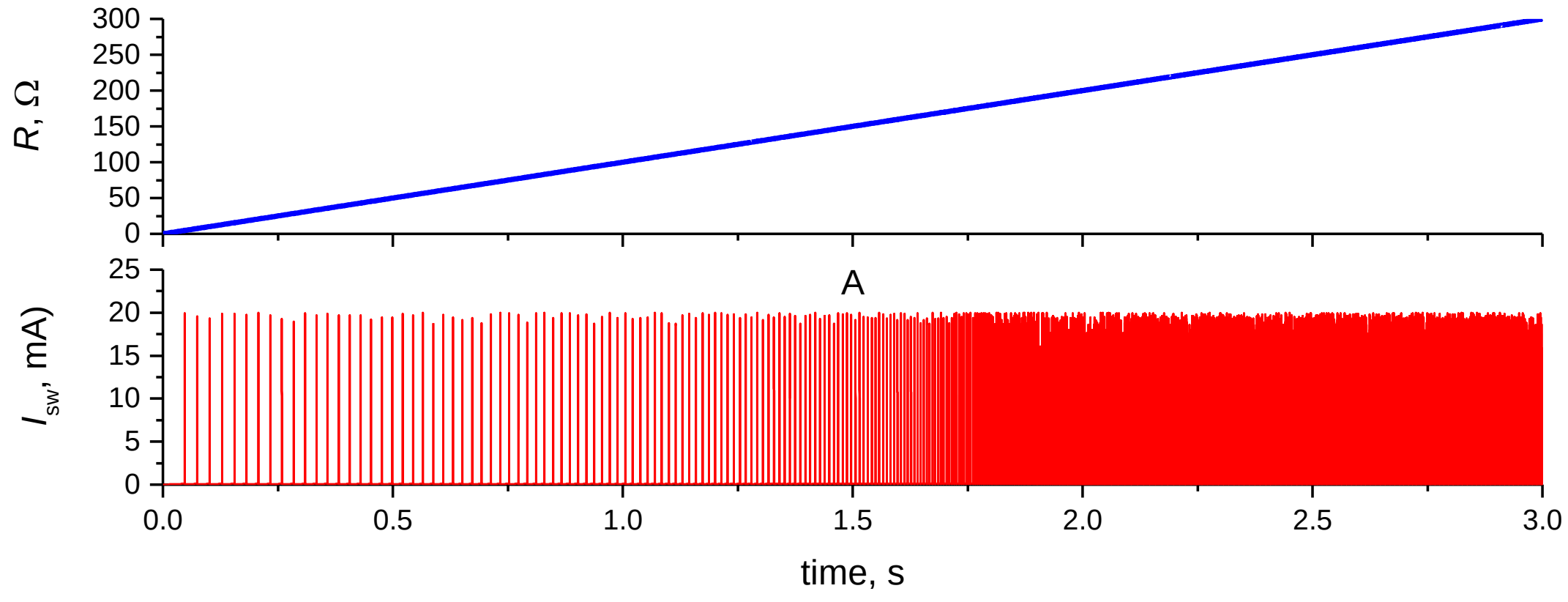

**Figure 4.** The current oscillogram through the switch, when the resistance $R(t)$ changes uniformly with time. The circuit parameters: $U_{th}$=4 V, $U_h$=2 V, $R_{off}$ =40 kΩ and $R_{on}$=200 Ω, $I_0$=150 μA, $C_1$=10 nF, $C_2$=1 μF, $R$=200 Ω. The simulation was performed in the LTspice software.

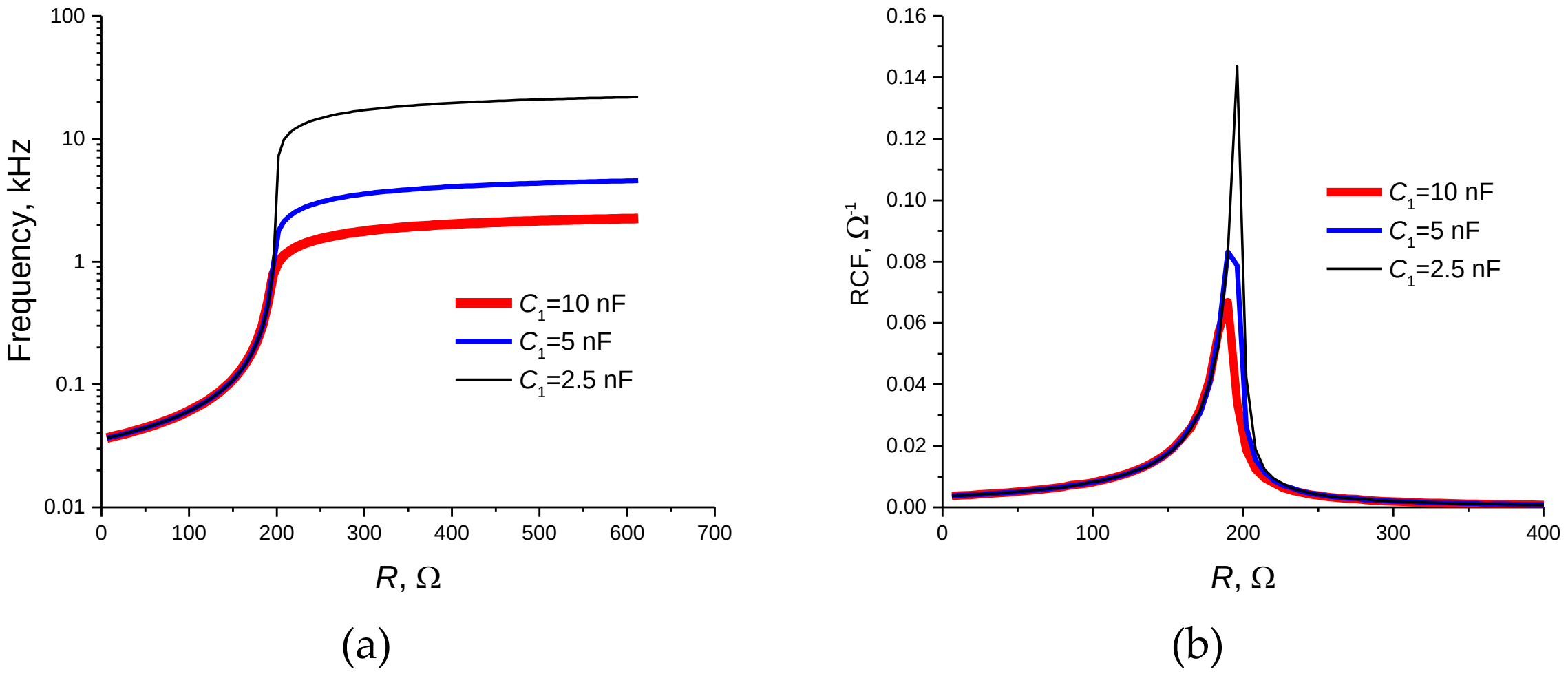

**Figure 5.** Dependences of the oscillation frequency of the circuit (a) and RCF (b) on the value of the control resistance $R$ for various capacitance values $C_1$. The simulation was performed according to the method (section 2) in the Mathcad software.

To assess the sensitivity of the circuit and to determine a change in the generation frequency to a change in resistance, we introduce the "resistive coefficient of frequency" (RCF) defined as:

$$\mathrm{RCF} = \frac{1}{F}\frac{\partial F}{\partial R} \quad (11).$$

Equation (11) expresses the relative change in frequency with varying resistance, and is introduced by analogy with the temperature coefficient of resistance TCR [24]. The RCF graph for various $C_1$ values is presented in Figure 5b. For all the curves, there is a sharp maximum at $R$=$R_{RCF}$ ~ 190 Ω. The maximum RCF≈0.14 corresponds to the smallest capacitance $C_1$ = 2.5 nF, and the greater is the value of RCF, the higher is the ratio $C_2/C_1$.

Therefore, the studied circuit shows the nonlinear dependence $F(R)$, with a maximum at the point $R$=$R_{RCF}$, and the highest sensitivity in this area.

The effect of a sharp transition from low-frequency to high-frequency oscillations can be used in neural circuits to model frequency coding with a sigmoid-like activation function. Figure 6 demonstrates the approximation of the function $F(R)$ for $C_1$ = 10 nF by a dependence:

$$F_{app}(R) = \frac{A_1 \cdot \left(1 - \exp(-A_3 R)\right)}{1 + \exp\left(-A_2 (R - A_4)\right)} + A_5 \quad (13),$$

where the coefficients $A_1$=2398.8 Hz, $A_2$=0.0848 Ω$^{-1}$, $A_3$=0.00415 Ω$^{-1}$, $A_4$=196 Ω, $A_5$=54 Hz.

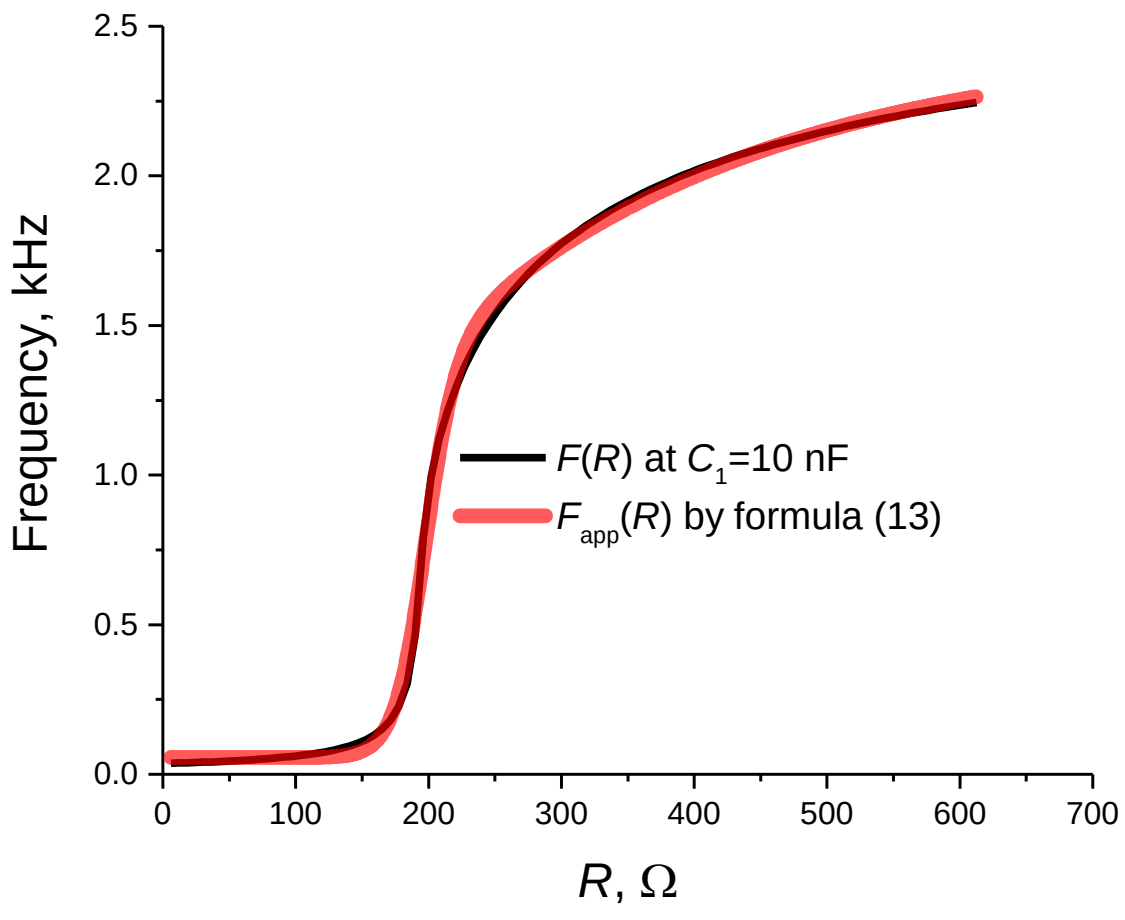

**Figure 6.** $F(R)$ dependence and its approximation by a sigmoid-like function (13). $A_1$=2398.8 Hz, $A_2$=0.0848 $\Omega^{-1}$, $A_3$=0.00415 $\Omega^{-1}$, $A_4$=196 Ω, $A_5$=54 Hz.

Approximation (13) represents an exponential sigmoid $[1+\exp(-A_2(R-A_4))]^{-1}$, multiplied by the growing function $A_1\cdot[1-\exp(-A_3R)]$. Coefficient $A_5$ corresponds to the minimum oscillation frequency, when $R$=0, and the dynamics of the circuit corresponds to oscillations of a single-capacitor circuit (see Figure 1b) at $C_0$=$C_2$. Coefficient $A_1$ corresponds to the maximum oscillation frequency $A_1$=$F_{max}$, when $R$ tends to infinity, and the dynamics of oscillations corresponds to a single-capacitor circuit (see Figure 1b), but at $C_0=(C_1\cdot C_2)/(C_1+C_2)$. The coefficient $A_4$ has a value close to the value of $R_{RCF}$ ($A_4$~$R_{RCF}$). The resistance value, inverse to the coefficient $A_2$, determines the width of maximum frequency increase $\Delta R=1/A_2$, and the resistance $R=1/A_3$ characterizes the transition point of the dependence $F(R)$ to saturation.

A detailed investigation of the circuit presented in the current study can be performed using LTspice files (see Supplementary Materials).

## 4. Discussion

The described circuit can be applied as in receptor modelling, as in modelling the interconnections of neurons in the network.

*Receptor modeling.*

To use the proposed circuit as an artificial receptor, it is necessary to convert the action force into a change in the electrical resistance of the sensing element. For example, resistive bend sensors, which are available on industrial scale, can be used to detect the bend. These sensors can be based on carbon powder nanoparticles, nanotubes, silver, copper, nickel, platinum, palladium and some conductive polymers [25].

To emulate temperature receptors, materials with a high temperature coefficient of resistance can be used as resistive sensors. Some metals, nickel, iron, tungsten [24], and transition metal oxides, for example, $VO_2$ [26], have a high TCR. In addition, not only a sensitive sensor can be made on $VO_2$ films, but also a switching element with an S-shaped I – V characteristic used in the circuit under study [27].

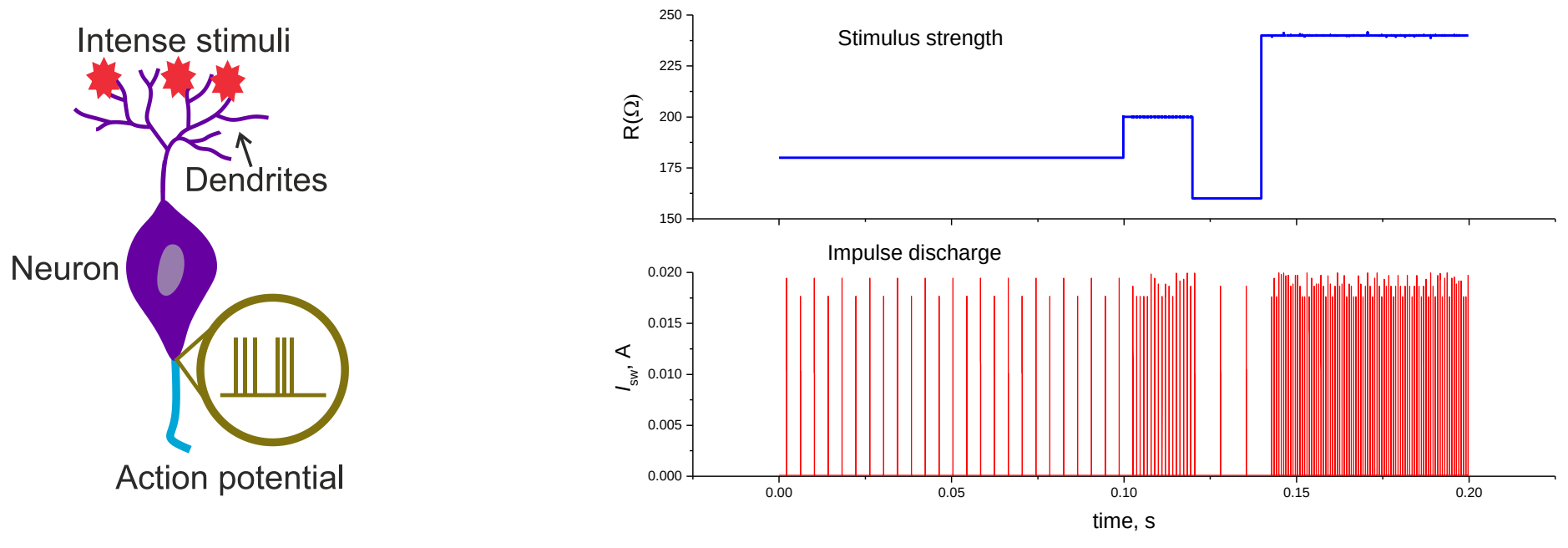

(a) (b)

**Figure 7.** Schematic representation of the neuron (a) and oscillogram of the current pulses when varying the control resistance of the circuit (b), demonstrating the relationship between the intensity of the action and the frequency of the output spikes.

Figure 7a presents a schematic representation of a neuron, and Figure 7b demonstrates an oscillogram of the current pulses of the switch when varying the resistance of the control resistor, which can be controlled by external stimuli, the magnitude of the bend, temperature, etc. As demonstrated in Figure 5, the advantage of our circuit is the high sensitivity of the output frequency to resistance change. For example, at $R$~$R_{RCF}$, a change in resistance by one ohm can lead to a change in frequency by 14%. The sensitivity of the circuit can be adjusted by the circuit parameters and the choice of the region of resistance change. All this extends the range of application of the circuit for converting the intensity of external influences into the frequency of output oscillations.

*Modelling of neurons.*

In the rate coding model, the frequency of generation of output spikes depends on the number of incoming neurons for a certain period of time. In order to create a network (Figure 8a) based on the circuit of the oscillator LIF neuron we proposed, it is necessary to convert the input frequencies into the value of the control resistance. Several options for frequency-resistance conversion modules (CM) can be offered.

1. Installing a pulse counter at the input, which resets its value to the memory register after a certain period. The value from the register goes to the digital-to-analog converter controlling the transistor resistance (Figure 8b). This circuit is difficult to implement; however, it can support several inputs of a neuron, including the inhibitor type, when the main pulse counter does not increase the count, but decreases it.
2. Using a storage capacitor, and its charge will increase with each pulse. The capacitor will be discharged through a shunt resistor, and the voltage from the capacitor is supplied to the transistor or phototransistor that sets the resistance *R* (Figure 8c).
3. The circuit in Figure 8d uses thermal effects and does not contain silicon components (transistors). Incoming spikes heat up the resistive element, and it, in turn, heats the heat-sensitive resistive sensor, which controls the circuit we have proposed. As the cooling of the heater is much slower than the time of heating it with short input spikes, the circuit will have the effect of integrating the input spikes [17] and gradual regulating the output resistance.

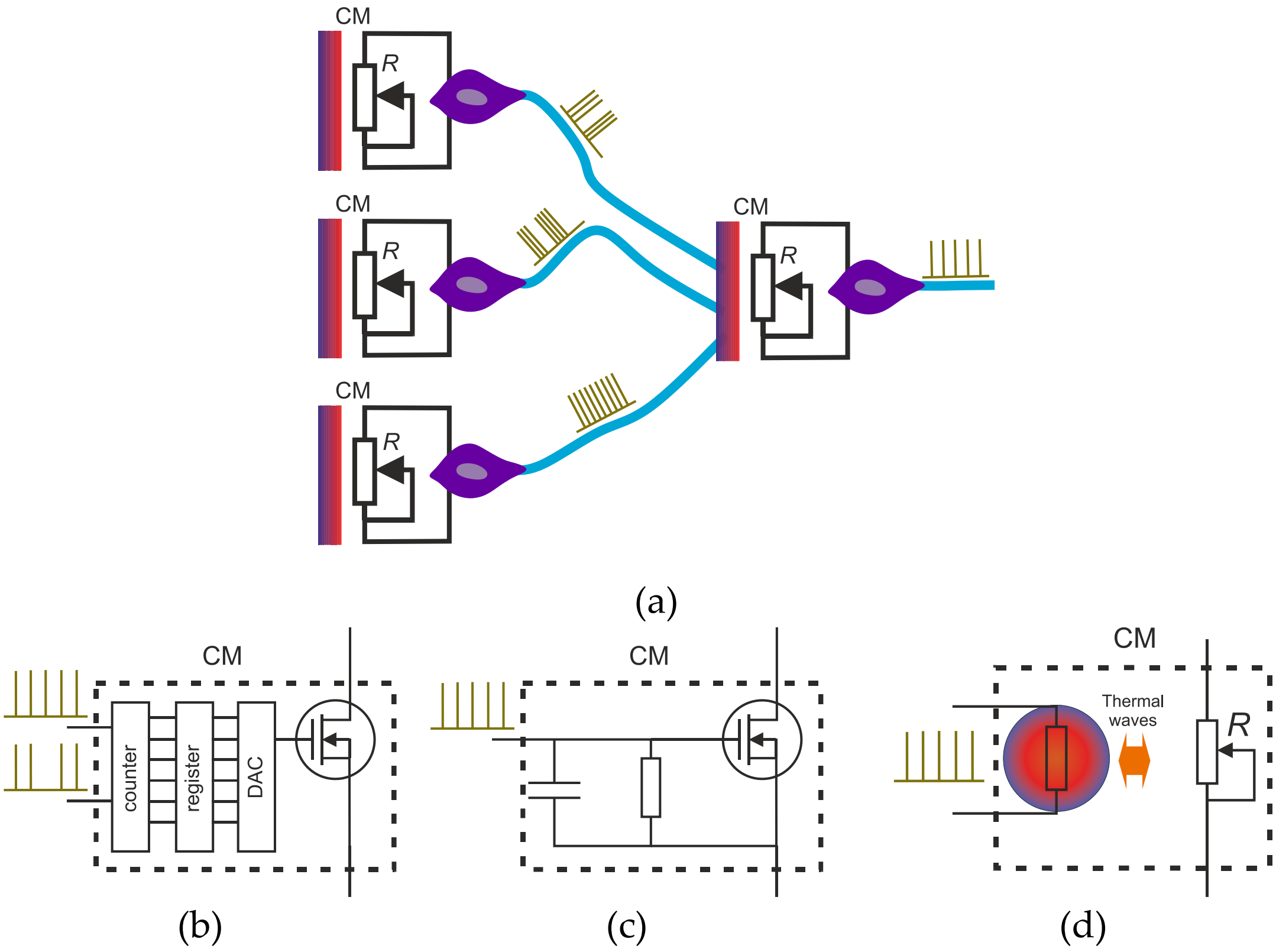

**Figure 8.** The circuit of a neural network based on an oscillator controlled by resistance and a frequency-resistance CM (a). Option of CM based on the counter and transistor (b). Option of CM based on an integrating capacitor and transistor (s). Option of CM based on thermal effect (d).

In addition, the literature describes the “Diff-Pair Integrator” circuit [28], which translates the oscillation frequency into a control current, however, it has a more complex operating principle and contains several silicon transistors.

In contrast to the circuit described in [16], where the frequency is controlled by the value of the supply current $I_0$, our circuit allows the implementation of a resistive-frequency control and can be interfaced with various resistive sensors. In addition, our circuit allows the implementation of a neural network with rate coding, and, among other options, without the use of silicon components (transistors).

## 5. Conclusions

In the current study, we proposed the oscillator circuit with a nonlinear dependence of the oscillation frequency on the control resistance. The circuit has a sigmoid-like activation function, with the effect of a sharp transition from low frequency to high frequency oscillations, which can be used to create neural networks with rate coding. Options of frequency-resistive interfaces are proposed, which allow converting the frequency of arrival spikes to the value of the control resistance.

We hope the results of this study would spark the interest of researchers in the experimental implementation of the proposed circuits, the mathematical analysis of dynamics of the systems with non-linear S-elements, and the development of oscillator and spike neural networks based on switching S-elements.

**Acknowledgments:** The authors express their gratitude to Dr. Andrei Rikkiev for the valuable comments in the course of the article translation and revision.